\documentclass{article}

\usepackage{microtype}
\usepackage{graphicx}
\usepackage{subfigure}
\usepackage{booktabs}
\usepackage{multirow}
\usepackage{array}
\usepackage{float}
\usepackage{hyperref}

\usepackage[accepted]{mlsys2025}

\mlsystitlerunning{Toward a Small ML Runtime Stack for Raspberry Pi 5 QPUs}

\begin{document}

\twocolumn[
\mlsystitle{Toward a Small ML Runtime Stack for Raspberry Pi 5 QPUs}

\mlsyssetsymbol{new}{$\dagger$}

\begin{mlsysauthorlist}
\mlsysauthor{Yiannis Hadjiyianni}{ntua}
\mlsysauthor{Panagiotis Michelakis}{ntua,new}
\mlsysauthor{Dimitrios Stamoulis}{hit}
\end{mlsysauthorlist}

\mlsysaffiliation{ntua}{School of Electrical and Computer Engineering, National Technical University of Athens, Athens, Greece}
\mlsysaffiliation{hit}{Department of Computer Science and Technology, Harbin Institute of Technology, Harbin, China}

\mlsyscorrespondingauthor{Yiannis Hadjiyianni}{yiannisha@synkrasis-labs.com}

\vskip 0.3in

\begin{abstract}
We present a QPU-\textit{first} ML runtime stack for Raspberry Pi 5's VideoCore VII QPU, built on top of the \texttt{py-videocore7} assembly library\textsuperscript{1}. The system comprises reusable tiled matrix-multiplication substrate, GEMM-backed convolution, a single-head attention-style core, persistent executors, and integer execution based on \texttt{smul24} instructions. For dense integer kernels, packed INT16-input with INT32 accumulation achieves nearly two orders of magnitude higher throughput over NumPy. Across operations (min/max, pooling, convolution, attention), we report improved performance over both PyTorch and NumPy. Our preliminary results indicate that Raspberry QPUs can serve as a practical execution substrate towards accelerating AI model execution at the edge.
\end{abstract}
]

\printAffiliationsAndNotice{\textsuperscript{$\dagger$}Currently with new affiliation.}

Raspberry Pi is widely deployed in edge and IoT systems~\cite{panopoulos2026athena}, yet its integrated QPU is rarely treated as an ML execution device. To our knowledge, there is no end-to-end runtime stack for QPUs: all existing inference engines (\textit{e.g.}, ONNX~\cite{onnx}, Ollama~\cite{ollama}) target its ARM CPU, leaving a gap between low-level QPU programmability and reusable inference-runtime support~\cite{ardakani2025llmpi, Wei_2025}. This undergraduate student-led work examines whether QPUs can support structured runtime execution.

\textbf{Methodology}. We build on \texttt{pyvideocore7}~\citep{pyvideocore7}, which provides Python-side \textit{assembly} generation, kernel loading, and dispatch for VideoCore VII QPUs. We implement \textbf{custom} QPU kernels, a tiled execution substrate, persistent executors, and an operator layer. The core design choice is to reuse a tiled matrix-multiplication substrate across multiple workloads. Convolution is implemented as GEMM-backed convolution through lowering, while the attention operator is a single-head dot-product attention core of the form $O=(QK^\top)V$. We build persistent executors that assemble kernels and allocate device buffers once; our cache uniforms and dispatch metadata for repeated execution. Moreover, we optimize for integer execution rather than using generic tensor-accelerator (Torch) semantics: we implement integer kernels that conform to VideoCore arithmetic constraints induced by the \texttt{smul24} instruction. These operand-range contracts propagate to higher-level operators to determine feasible tiling, packing, and accumulation at runtime. Our evaluation therefore reports both operand-level throughput and operator-level runtime, while separating one-time setup, cached steady-state execution (QPU-C), and execute-only (QPU-E) kernel time.

\begin{table*}[t!]
\vspace{-10pt}
\caption{Integer GEMM throughput (GOPS).}
\label{tab:gemm}
\vskip 0.01in
\centering
\scriptsize
\begin{tabular}{lccc|ccc}
\toprule
\multirow{4}{*}{\textbf{Size}} & \multicolumn{3}{c|}{
\textbf{INT32}} & \multicolumn{3}{c}{\textbf{Packed}} \\
& \multicolumn{3}{c|}{\textbf{GEMM}} & \multicolumn{3}{c}{\textbf{INT16-Input/INT32-Accumulation}} \\
\cmidrule(lr){2-4}\cmidrule(lr){5-7}
& \textbf{NumPy} & \textbf{QPU} & \textbf{Speedup} & \textbf{NumPy} & \textbf{QPU} & \textbf{Speedup} \\
\midrule
256  & 1.18 & 6.30  & 5.34$\times$  & 1.17 & 6.30  & 5.40$\times$  \\
512  & 0.57 & 16.69 & 29.49$\times$ & 0.57 & 16.82 & 29.70$\times$ \\
768  & 0.57 & 20.49 & 36.25$\times$ & 0.60 & 20.49 & 33.91$\times$ \\
1024 & 0.21 & 10.73 & 50.15$\times$ & 0.23 & 21.67 & 94.38$\times$ \\
\bottomrule
\end{tabular}%
\end{table*}

\begin{table*}[t!]
\caption{Operator-level Raspberry PI performance. }
\label{tab:ops}
\vskip 0.01in
\centering
\scriptsize
\begin{tabular}{lccccc}
\toprule
\multirow{2}{*}{\textbf{Operator}$^{\P}$} & \multirow{2}{*}{\textbf{Setting}} & \multirow{2}{*}{\textbf{NumPy}} & \multirow{2}{*}{\textbf{PyTorch}} & \textbf{QPU-C}$^{\ast}$  & \textbf{QPU-E} \\
 & & & & \textbf{Steady-State (Cached)} & \textbf{Execute-Only} \\
\midrule
Min/max & INT32, 12 cores & 6.27 & 6.17 & \textbf{6.50} & --- \\
Min/max & INT16, 12 cores & \textbf{6.71} & 6.26 & 6.02 & --- \\
AvgPool {\tiny 2$\times$2 / 2} & INT32, 12 cores & 0.72 & 0.69 & \textbf{2.62} & --- \\
MaxPool {\tiny 2$\times$2 / 2} & FP32, 12 cores & 2.66 & \textbf{3.03} & 2.65 & --- \\
Conv2D & FP32 & \textbf{21.69} & 12.74 & 16.25 & 19.87 \\
Conv2D & INT32 & 1.44 & 8.30 & \textbf{15.08} & 18.60 \\
Conv2D & INT16-IN/INT32-ACC & 1.45 & 15.45$^{\dagger}$ & \textbf{15.97} & 18.43 \\
Attention  & Core total, FP32 & \textbf{82.35} & 5.97$^{\ddagger}$ & 12.74 & 14.31 \\
Attention & Core total, INT32 & 1.13 & 1.79 & \textbf{12.81} & 14.41 \\
\bottomrule
\end{tabular}%
\vskip 0.04in
\scriptsize
$^{\P}$ Measured as min/max, pooling: GiB/s; convolution, attention: GOPS. \\
$^{\ast}$ Cached: repeated execution w/ steady-state kernels and buffers
w/o one-time setup. \\
$^{\dagger}$ Native PyTorch INT16 convolution is used as a speed baseline only on this build. \\
$^{\ddagger}$ The FP32 attention baseline is native SDPA and softmax with scale $1.0$. \\
\end{table*}

\textbf{Results}. Table~\ref{tab:gemm} summarizes dense integer GEMM performance. Standard INT32 GEMM peaks at 20.49~GOPS, while packed INT16-input with INT32 accumulation reaches 21.67~GOPS, with up to 94.38$\times$ speedup over NumPy. Table~\ref{tab:ops} reports operator-level performance. Integer operators show strongest gains, \textit{e.g.}, INT32 avgpool reaches 2.62~GiB/s above both NumPy (0.72~GiB/s) and PyTorch (0.69~GiB/s). INT32 GEMM-backed convolution reaches 15.08 and 18.60~GOPS for QPU-C and QPU-E, respectively. For the INT32 attention, throughput is 12.81~GOPS for QPU-C \textit{vs}. 14.41~GOPS for QPU-E. The gap between cached and execute-only is small once setup is amortized, supporting a runtime design with persistent executors and precompiled kernels. \textbf{CNN performance:} In preliminary experimentation, we have successfully run an end-to-end QPU-first \textit{LeNet} model, where the 12-core execute-only INT32 pipeline reaches 4.08 GOPS, compared to 0.83 GOPS for NumPy, a nearly 5$\times$ performance increase.

\textbf{Discussion}. This work-in-progress investigation established a technically meaningful execution substrate on VideoCore VII: a reusable tiled backbone, operator implementations that inherit its layout and dispatch strategy, explicit integer contracts tied to the underlying machine, and a measurement methodology that distinguishes compilation, host orchestration, and steady-state device execution. In our ongoing work~\cite{michelakis2025core}, we aim towards a full production backend, with broader operator coverage, CPU--QPU scheduling, and lightweight end-to-end LLM model inference on Raspberry Pi.

\bibliography{main}
\bibliographystyle{mlsys2025}

\end{document}